\documentstyle[12pt]{article}
\newcommand{\simleq}
{\mbox{\raisebox{-0.5ex}{$\textstyle \sim$}
 \raisebox{ 0.8ex}{$\textstyle  \!\!\!\!\!\! <$  }}}

\newcommand{\AmS}{{\protect\the\textfont2
  A\kern-.1667em\lower.5ex\hbox{M}\kern-.125emS}}

\begin{document}
\title{Gravitational phase transition of heavy neutrino matter}
\author{}
\maketitle
\begin{center}
Neven Bili\'{c} \footnote{permanent address: Rudjer
Bo\v skovi\'{c} Institute, 10000 Zagreb, Croatia}, David Tsiklauri
and Raoul D.\ Viollier \\Department of Physics, University of Cape
Town\\ Rondebosch 7701, South Africa
\end{center}

\abstract{In the framework of the Thomas-Fermi model at finite temperature, we show that
a cooling nondegenerate gas of massive neutrinos will, at a certain temperature,
undergo a phase transition in which quasi-degenerate supermassive neutrino stars
are formed through gravitational collapse.  For neutrinos in the mass range of 10
to 25 keV/c$^{2}$, these compact dark objects could mimic the role of supermassive
black holes that are usually assumed to exist at the centres of galaxies and
quasi-stellar objects. Astrophysical implications and constraints
on the neutrino mass are discussed for this scenario.}

\section{Introduction}
One of the most tantalizing puzzles of this Universe is the dark
matter problem, whose existence is inferred from the observed
flat rotation curves in spiral galaxies \cite{Fab,Trim}, the
diffuse emission of X-rays in elliptical galaxies and clusters of
galaxies, as well as from cluster dynamics.  The discrepancy
between baryon abundances derived from standard
nucleosynthesis calculations and the observed amount of luminous
matter suggests that most of the
baryonic matter in this Universe is nonluminous, and such an
amount of dark matter falls suspiciously close to that required
by the galactic rotation curves.  However, although a significant
component of the dark matter in galactic halos may be baryonic
\cite{Copi}, the dominant part of dark matter in this Universe is believed to
be nonbaryonic, such as to make this Universe nearly critical.  Many candidates have been proposed \cite{Peeb},
both baryonic as well as nonbaryonic, to explain the dark matter
problem, but the issue of the nature of dark matter is still far
from being resolved.

In this paper, we will focus on nonbaryonic dark matter
consisting of weakly interacting heavy fermions.  We are
particularly interested in fermions with masses between 10 and
25 keV/$c^2$ which cosmologically fall into the category of warm dark
matter.  The reason for choosing fermions in this mass range is
that these could form supermassive degenerate objects which may
explain, without invoking the black hole hypothesis,
some of the features observed around the supermassive compact
dark objects of masses ranging from 10$^{6.5} M_\odot$ to 10$^{9.5} M_\odot$
which have been reported at the centers of a number of galaxies
[5-10] including our own [11-15] and quasi-stellar
objects [16-19,21].

Interpreting this fermion as a 10 to 25 keV/$c^2\;\tau$ neutrino,
is neither in conflict with
particle and nuclear physics nor with astrophysical observations
\cite{RDV8}.  On the contrary, if the conclusion of the LSND
collaboration, which claims to have detected $\bar{\nu}_{\mu}
\rightarrow \bar{\nu}_{e}$ flavour oscillations \cite{Ath22}, is
confirmed, and the quadratic see-saw mechanism involving the up,
charm and top quarks \cite{Gel23,Yan24}, is the correct mechanism
for neutrino mass generation, the $\nu_{\tau}$ mass may well be
between 6 and 32 keV/$c^2$ \cite{Bil34} which falls into the
cosmologically forbidden range between 93 $h^{-2}$ eV/$c^2$ and 4 GeV/$c^2$
with $0.4\;\;\simleq h\;\;\simleq 1$.  It is well known that
such a quasistable neutrino would pose serious problems in
cosmology, as it would generate an early matter
dominated phase, starting perhaps as early as a couple of weeks
after the Big-Bang \cite{Kol21,RDV8,Lur5}.  As a direct consequence of
the existence of such a heavy neutrino, a critical Universe would
have reached the current microwave
background temperature in less than 1 Gyr, i.e.\ much too
early to accommodate the oldest stars in globular clusters,
cosmochronology and the Hubble expansion age \cite{Kol21}.

It is conceivable, however, that, in the presence of such heavy
neutrinos, the early Universe might have evolved quite
differently than described in the Standard Model of Cosmology.
In particular, primordial neutrino stars might have been formed
in local condensation processes during a gravitational phase
transition, shortly after the neutrino matter dominated epoch
began.  The latent heat produced in such a first-order phase
transition, apart from reheating the gaseous neutrino phase,
might have contributed towards reheating the radiation background as
well. Moreover, the bulk part of the heavy neutrinos and
antineutrinos would have annihilated into light neutrinos and
antineutrinos via the $Z^0$ in the dense interior of the
supermassive neutrino stars \cite{RDV8,Lur5,RDV7}.  In this
context, it is interesting to note that, even within the
(homogeneous) Standard Model of Cosmology, the reason why
neutrino masses larger than 4 GeV/$c^2$ are again cosmologically
allowed, is precisely because annihilation of the heavy
neutrinos reduces the matter content of the present
Universe.  Since reheating and annihilation will increase the age
of the Universe, i.e.\ postpone
the time when the Universe reaches the current microwave
background temperature \cite{Kol21,RDV8}, it does not seem
excluded that a quasistable massive neutrino in the mass range
between 10 and 25 keV is compatible with cosmological
observations \cite{RDV8,Lur5}.

In fact, following ref. \cite{Bil34,Bil25}, we will show
in this paper that degenerate neutrino stars
may indeed have been formed during a
gravitational phase transition in the early Universe.  While the
existence of such a first-order phase transition is well
established in the framework of the Thomas-Fermi model at finite
temperature \cite{Bil25}, the mechanism through which the latent
heat is released during the phase transition and dissipated into
observable and perhaps unobservable matter or radiation remains
to be identified. At this stage, however, it is still not clear
whether an efficient dissipation mechanism can be found within
the minimal extension of the Standard Model of Particle Physics
or whether new physics is required in the right-handed neutrino
sector, in order to avoid supercooling of heavy neutrino matter.
Thus, we will henceforth assume that an efficient dissipation
mechanism exists, within or beyond the Standard Model of Particle
Physics, in order to make sure that neutrinos and antineutrinos
can settle in the state of lowest energy in a reasonable period
of time.

\section{Cold neutrino stars and astrophysical implications}
The ground state of a gravitationally condensed neutrino cloud,
with mass below the Oppenheimer-Volkoff (OV) limit, is a cold
neutrino star \cite{RDV8,Lur5,RDV7,RDV6} , in which the degeneracy pressure balances
the gravitational attraction of the neutrinos.  Inserting the
relativistic equation of state for degenerate neutrino matter into the
general relativistic Tolman-Oppenheimer-Volkoff equations of
hydrostatic equilibrium, the upper
bound for the mass of such a neutrino star, the OV
limit, turns out to be  $M_{OV} = 0.54195 \;\; m^{3}_{p \ell} m_{\nu}^{-2}
g_{\nu}^{-1/2}$ \cite{Mun40}.  Here $m_{p \ell} =
(\hbar c/G)^{1/2} = 1.22205 \cdot 10^{19}$ GeV/$c^{2}$ is the
Planck mass and $g_{\nu}$ the spin degeneracy factor of the
neutrinos and antineutrinos, i.e. $g_{\nu} = 2$ for Majorana or
$g_{\nu} = 4$ for Dirac neutrinos and antineutrinos.  The radius
of such a compact dark object is $R_{OV} = 4.4466
R^{s}_{OV}$, where $R_{OV}^{s} = 2 GM_{OV}/c^{2}$ is the
Schwarzschild radius of the mass $M_{OV}$.  There
is thus little difference between a neutrino star at the OV
limit and a supermassive black hole of the same mass, a few
Schwarzschild radii away from the object, in particular, since
the last stable orbit around a black hole has also a radius of
$3R_{OV}^{s}$.  For nonrelativistic neutrino stars that are well
below the OV limit, mass and radius scale as $MR^{3} = 91.869
\;\; \hbar^{6} G^{-3} m_{\nu}^{-8} (2/g_{\nu})^{2}$ \cite{RDV8}, as
follows directly from the nonrelativistic Lan\'{e}-Emden equation
with polytrope index $3/2$, subject to the usual boundary conditions at
the center and the surface of the neutrino star.  If we
generalize the boundary condition to include a point
mass at the center, the Lan\'{e}-Emden equation is able to describe
a degenerate neutrino halo around a baryonic star
\cite{RDV8,RDV6} as well.

The idea that some of the galactic nuclei
are powered by matter accretion onto supermassive black holes is based
on general theoretical arguments  and the observation of rapid time
variability of the emitted radiation
which implies relativistic compactness of the radiating object.
However, so far, there is no compelling evidence that supermassive
black holes actually do exist, as the spatial resolution of
current observations
is larger than $10^5$ Schwarzschild radii.
The standard routine in investigating the nature of the
dark mass distribution
at the centers of active galaxies, is to observe
stellar and gas dynamics. However, gas dynamics is usually regarded
as less conclusive, since it is responsive to nongravitational
forces such as magnetic fields.

Because of its relative proximity,
the best place to test the supermassive black hole paradigm is
perhaps the compact radio source Sgr A$^*$, located at the center of our galaxy.
Observations of
stellar motions near the galactic center
\cite{Eck26,Gen27}
and the low proper motion ($\leq$ 20 km sec$^{-1}$)
\cite{Bac35} of Sgr A$^*$ indicate, on the one hand,
that Sgr A$^*$ is a massive $M = (2.5 \pm 0.4) \times 10^6 M_\odot$
object that dominates the gravitational potential in the
inner ($\leq 0.5$ pc)
region of the galaxy. On the other hand,
observations of stellar winds and other gas flows in the vicinity
of Sgr A$^*$
suggest that the mass accretion rate is about $\dot M = 6 \times
10^{-6}M_\odot$yr$^{-1}$ \cite{Gen36}.  Thus, if Sgr A$^*$ is
indeed a supermassive
black hole, as many believe, the luminosity
of this object should be more than $10^{40}$ erg sec$^{-1}$, provided
the radiative efficiency is the customary 10\%.
However, observations indicate that the
bolometric luminosity of Sgr A$^\ast$ is actually less than $10^{37}$ erg
sec$^{-1}$.  This discrepancy, known as the so-called `blackness
problem' of the center of our galaxy, has been the source of
exhaustive debate in the recent past, in which the notion of a
`supermassive black hole on starvation' has been introduced.  A
much simpler solution to this problem could be to give up the
idea that Sgr A$^\ast$ is a black hole.

Indeed, an alternative model
for the mass distribution at the galactic
center was proposed \cite{Tsi38} recently, in which the customary supermassive
black hole is replaced by
a supermassive neutrino star.  It was shown that a degenerate neutrino star
of mass $M = 2.5 \times 10^6 M_\odot$, consisting of neutrinos
with mass $m_\nu \geq 12.0$ keV$/c^2$ for $g_{\nu} = 4$, or $m_\nu \geq
14.3$ keV$/c^2$ for $g_{\nu} =2$, does not contradict the
current observational data.  This can be seen in Fig.1, where the
mass enclosed within a given distance from Sgr A$^*$ is plotted
for a $M = 2.5 \times 10^{6} M_{\odot}$ neutrino star
embedded in a standard stellar cluster with a central density
$\rho_{0} = 4 \times 10^6 M_\odot /{\rm pc}^3$ and a core radius
$R_{\rm core}=0.38$ pc. We thus conclude that for certain values
of $m_{\nu}$ and $g_{\nu}$, this model of Sgr A$^*$ is consistent
with the mass distribution data.
\begin{figure}[p]
%\vspace{12cm}
\caption{Various models for the mass enclosed within a distance
from Sgr A$^*$: (i) a $M = 2.5 \times 10^{6} M_{\odot}$ neutrino
star plus stellar cluster, with neutrino masses in the range
of 10-25 keV$/c^2$ and degeneracy factors
$g_{\nu} = 2$ and $g_{\nu} = 4$, (ii) a $M = 2.5 \times 10^6
M_\odot$ black hole plus stellar cluster, (iii) a stellar
cluster only. Note that the curves for $m_\nu = 12.013$ keV$/c^2$
and $g_{\nu} = 4$, and $m_\nu = 14.285$ keV$/c^2$ and $g_{\nu} =
2$, coincide. Data points are taken from [13,14] and
references therein.}
\end{figure}

More recently, also the spectrum emitted by Sgr A$^*$ was calculated
in the framework of
standard accretion disk theory, assuming that
Sgr A$^*$ is a neutrino
star with radius $R = 30.3$ ld ($\approx 10^{5}$ Schwarzschild
radii), corresponding to a neutrino mass of $m_{\nu} = 12.0$
keV/$c^{2}$ for $g_{\nu} = 4$, or $m_{\nu} = 14.3$ keV for
$g_{\nu} = 2$ \cite{Tsi39}.  In such a scenario, the accretion
disk is contained entirely within the neutrino star.  The
calculated product of luminosity and frequency $\nu L_{\nu}$ is
presented in Fig.2.  The thick solid line represents the case of
a neutrino star with $\dot{M}/\dot{M}_{Edd} = 4 \times 10^{-3}$,
whereas thin solid line corresponds to $\dot{M} / \dot{M}_{Edd} =
10^{-4}$. Here $\dot{M}_{Edd} = 2.21 \cdot 10^{-8} M$ yr$^{-1}$
denotes the Eddington limit accretion rate.
The short-dashed line
describes a calculation with a
$M = 2.5 \times 10^6 M_\odot$  black hole,
$\dot{M}/\dot{M}_{Edd} = 10^{-4}$ and an accretion disk
extending from 3 to $10^5$ Schwarzschild radii. The long-dashed
line represents the case where the accretion rate is
artificially reduced to $\dot{M}/\dot{M}_{Edd} = 10^{-9}$.
As seen in Fig.2, the neutrino star model reproduces the observed spectrum from the radio to the near
infrared band very well.  Thus, as our model fulfils two of the
most stringent conditions, i.e.\ the mass distribution
\cite{Eck26,Gen27} and the emitted spectrum \cite{Nar37}, we can
conclude that the neutrino star scenario is consistent with the
currently available observational data.
\begin{figure}[p]
%\vspace{12cm}
\caption{Comparison of the theoretical and observed spectra of Sgr A$^*$.
The thick solid line corresponds to the case of a neutrino star
of mass $M = 2.5 \times 10^6 M_\odot$ with
$\dot{M}/\dot{M}_{Edd} = 4 \times 10^{-3}$, while the
thin solid line represents $\dot{M}/\dot{M}_{Edd} = 10^{-4}$. The
short-dashed line describes a  calculation for a $M = 2.5 \times
10^6 M_\odot$ black hole, with $\dot{M}/\dot{M}_{Edd} = 10^{-4}$
and an accretion disk extending from 3 to $10^5$ Schwarzschild
radii. The long-dashed line corresponds to the case when
$\dot{M}/\dot{M}_{Edd}$ is artificially reduced to $10^{-9}$.
Data points in the $<40$ GHz region are upper bounds
[35]. Note, that the thick solid line fits the most reliable data points
with error bars.}
\end{figure}

We now turn to the discussion of the constraints that our model
puts on the neutrino mass. Assuming that the most massive known compact
dark object, which has a mass $M = (3 \pm 1) \times 10^{9}
M_{\odot}$ and is located at the center of M87, is a
degenerate neutrino star at the OV limit, where it is close to being
a black hole, the neutrino mass must be in the range of 12.5 keV/$c^{2}
\leq m_{\nu} \leq 17.7$ keV/$c^{2}$ for $g_{\nu} = 2$, and 10.5
keV/$c^{2} \leq m_{\nu} \leq 14.9$ keV/$c^{2}$ for $g_{\nu} =
4$ (Fig.3).

If we interpret Sgr A$^\ast$ at the center of our galaxy as a
degenerate neutrino star with mass $M = 2.5 \times 10^{6}
M_{\odot}$ and a radius $R \leq 30.3$ ld, as derived in the
analysis by Genzel et al.\ \cite{Eck26,Gen27} shown in Fig.1, the
neutrino mass must be $m_{\nu} \geq 14.3$ keV/$c^{2}$ for $g_{\nu} =
2$, and $m_{\nu} \geq 12.0$ keV/$c^{2}$ for $g_{\nu} = 4$ (Fig.3).
Of course, for neutrino masses close to the minimal mass, the neutrino
star will be far away from being a black hole.  In fact in our
scenario, the much shallower gravitational potential governing the
accretion disk inside the neutrino star was essential for the
success of the calculation of the emitted spectrum of Sgr A$^*$.
The near black hole situation is obtained in the OV limit with
$m_{\nu} = 444$ keV/$c^{2}$ for $g_{\nu} = 2$, or $m_{\nu} = 374$
keV/$c^{2}$ for $g_{\nu} = 4$.
\begin{figure}[p]
%\vspace{14cm} %13
\caption{Neutrino mass limits based on various assumptions on the
supermassive compact dark objects at the center of M87 and our
galaxy.  Also shown are mass limits derived from a possible
degenerate neutrino halo around the sun. The dotted $\nu_{\tau}$
mass range is derived from the LSND experiment and the quadratic
see-saw mechanism.}
\end{figure}

Heavy neutrinos can of course also cluster around ordinary stars.
Let us therefore assume that the sun is surrounded by a
degenerate neutrino halo \cite{RDV8,RDV6}.  From the Pioneer 10
and 11 and Voyager 1 and 2 ranging data \cite{And41} we know that
the dark mass contained within Jupiter's orbit is $M_{d} = (0.12
\pm 0.027) \times 10^{-6} M_{\odot}$ and within Neptune's orbit
$M_{d} \leq 3 \times 10^{-6} M_{\odot}$.  Of course the Jupiter
data should be taken with caution, as Jupiter tends to eject
almost any matter within its orbit \cite{And41}.  Nevertheless,
taking Jupiter's data at face value, and interpreting the dark
matter as degenerate neutrino matter \cite{RDV8}, the neutrino mass limits
are 12.6 keV/$c^{2} \leq
m_{\nu} \leq 14.2$ keV/$c^{2}$ for $g_{\nu} = 2$ and 10.6
keV/$c^{2} \leq m_{\nu} \leq 12.0$ keV/$c^{2}$ for $g_{\nu} = 4$.
If the dark matter contained within Neptune's orbit is
interpreted as neutrino matter \cite{RDV8}, the mass limits are $m_{\nu} \leq
15.6$ keV/$c^{2}$ for $g_{\nu} = 2$, and $m_{\nu} \leq 13.1$
keV/$c^{2}$ for $g_{\nu} = 4$.  In summary, a mass range of 14.1 keV/$c^{2}
\leq m_{\nu} \leq 15.6$ keV/$c^{2}$ for $g_{\nu} = 2$, and 11.8
keV/$c^{2} \leq m_{\nu} \leq 13.1$ keV/$c^{2}$ for $g_{\nu} = 4$
seems to be consistent with all reliable data.

\section{Neutrino matter at finite temperature and the phase
transition}
A gas of massive fermions interacting only gravitationally has
interesting thermal properties that may
have important consequences in
astrophysics.  The canonical and grand-canonical ensembles for such a
system have been shown to have a nontrivial thermodynamical
limit \cite{Thi1,Her2}.  Under certain conditions these systems
will undergo a phase transition that is accompanied by
gravitational collapse \cite{Mes3}.  It is interesting to
note that this phase transition occurs only in the case of the attractive
gravitational interaction of neutral particles obeying Fermi-Dirac
statistics: it neither happens in the case of the repulsive
Coulomb interaction of charged fermions \cite{Fey4}, nor does it in the
case of the attractive gravitational interaction, when the
particles obey Bose-Einstein or Maxwell-Boltzmann statistics.  To
be specific, we henceforth assume that this neutral fermion is
the $\nu_{\tau}$ which is presumably the heaviest neutrino,
although this is not essential for most of the subsequent
discussion.  We, moreover, assume that the weak interaction can
be neglected.

The purpose of this chapter is to study the formation of a neutrino
star during a gravitational phase transition.  For simplicity,
let us assume that the neutrino star will be sufficiently below the
OV limit, so that we can treat this process
nonrelativistically \cite{Bil25}.  The gravitational potential energy $V(r)$
satisfies Poisson's equation
\begin{equation}
\Delta V   =   4 \pi G m_{\nu}^{2} n_{\nu},
\end{equation}
where the number density of the $\tau$ neutrinos (including antineutrinos) of
mass $m_{\nu}$ can be expressed in terms of the Fermi-Dirac distribution
at a finite temperature $T$ as
\begin{equation}
n_{\nu} (r)   =   \frac{g_{\nu}}{4 \pi^{2} \hbar^{3}} (2 m_{\nu} kT)^{3/2}
I_{\frac{1}{2}} \left( \frac{\mu - V(r)}{kT} \right).
\end{equation}
Here $I_{n} (\eta)$ is the Fermi function
\begin{equation}
I_{n} (\eta)   =   \int^{\infty}_{0} \frac{\xi^{n} d \xi}{1 + e^{\xi -
\eta}},
\end{equation}
and $\mu$ the chemical potential.  It is convenient to introduce
the normalized reduced potential
\begin{equation}
v   =   \frac{r}{m_{\nu} GM_{\odot}} ( \mu - V),
\end{equation}
$M_{\odot}$ being the solar mass, and the dimensionless variable $x =
r/R_{0}$ with the scale factor
\begin{equation}
R_{0} = \left( \frac{3 \pi \hbar^{3}}{4 \sqrt{2} m^{4}_{\nu} g_{\nu} G^{3/2}
M_{\odot}^{1/2}} \right)^{2/3} = 2.1377\;\;{\rm lyr} \left(
\frac{17.2\;\;{\rm keV}}{ m_{\nu} c^{2}} \right)^{8/3} g_{\nu}^{-2/3}.
\end{equation}
Equation (1) then takes the simple form
\begin{equation}
\frac{1}{x} \frac{d^{2} v}{d x^{2}} = - \frac{3}{2} \beta^{-3/2}
I_{\frac{1}{2}} \left( \beta \frac{v}{x} \right),
\end{equation}
\noindent
where we have introduced the normalized inverse temperature $\beta =
T_{0}/T$, with $T_{0} = m_{\nu} GM_{\odot}/kR_{0}$.  In equation (6) we
recover, at zero temperature, the well-known Lan\'{e}-Emden differential
equation \cite{RDV6,RDV8}
\begin{equation}
\frac{d^{2} v}{dx^{2}} = - \frac{v^{3/2}}{\sqrt{x}}.
\end{equation}

The solution of the differential equation (6) requires boundary
conditions.  We assume here that the neutrino gas is enclosed in a
spherical cavity of radius $R$ corresponding to $x_{1} =
R/R_{0}$, in order to prevent the gas from escaping to infinity.  We
further require the total neutrino mass to be $M_{\nu}$, and we
allow for the possibility of a pointlike mass $M_{B}$ at the origin,
which could be, e.g., a compact seed of baryonic or any other
compact matter.  $v(x)$ is
then related to its derivative at $x = x_{1}$ by
\begin{equation}
v' (x_{1}) = \frac{1}{x_{1}} \left( v(x_{1}) - \frac{M_{B} +
M_{\nu}}{M_{\odot}} \right),
\end{equation}
which, in turn, is related to the chemical potential by $\mu = k T_{0} v'
(x_{1})$.  $v(x)$ at $x=0$ is given by the point mass at the
origin, i.e.\ $M_{B}/M_{\odot} = v(0)$.

Similarly to the case of the Lan\'{e}-Emden equation, it is easy to show
that equation (6) has a scaling property: if $v(x)$ is a solution of
equation (6) at a temperature $T$ and a cavity radius $R$, then $\tilde{v}
(x) = A^{3}v(Ax)$, with $A > 0$, is also a solution at the
temperature $\tilde{T} = A^{4}T$ and the cavity radius $\tilde{R} = R/A$.

It is important to note that only those solutions that minimize the
free energy are physical.  The free-energy functional is defined
as \cite{Her2},
\begin{equation}
F[n_{\nu}] = \mu[n_{\nu}] N_{\nu} - W[n_{\nu}] - kTg_{\nu} \int \frac{d^{3} rd^{3} p}{(2 \pi
\hbar)^{3}} {\rm ln} \left[ 1 + {\rm exp} \left( -
\frac{p^{2}}{2 m_{\nu} kT} - \frac{V[n_{\nu}]}{kT} +
\frac{\mu[n_{\nu}]}{kT} \right) \right],
\end{equation}
where
\begin{equation}
V[n_{\nu}] = -Gm^{2}_{\nu} \int d^{3} r' \frac{n_{\nu}(r')}{|\vec{r} - \vec{r}'|},
\end{equation}
and
\begin{equation}
W[n_{\nu}] = - \frac{1}{2} Gm_{\nu}^{2} \int d^{3} rd^{3} r'
\frac{n_{\nu}(r) n_{\nu}(r')}{|\vec{r} - \vec{r}'|}.
\end{equation}
The chemical potential in equation (9) varies with density, so
that the number of neutrinos $N_{\nu} = M_{\nu}/m_{\nu}$ is kept
fixed.

All the relevant thermodynamical quantities, such as number
density, pressure, free energy, energy and entropy, can be
expressed in terms of $v/x$, i.e.
\begin{equation}
n_{\nu} (x) = \frac{M_{\odot}}{m_{\nu} R_{0}^{3}} \frac{3}{8 \pi}
\beta^{-3/2} I_{\frac{1}{2}} \left( \beta \frac{v}{x} \right),
\end{equation}

\begin{equation}
P_{\nu}(x) = \frac{M_{\odot} T_{0}}{m_{\nu} R_{0}^{3} 4 \pi}
\beta^{-5/2} I_{\frac{3}{2}} \left( \beta \frac{v}{x} \right) =
\frac{2}{3} \epsilon_{{\rm kin}} (x),
\end{equation}

\begin{equation}
F  =  \frac{1}{2} \mu N_{\nu} + \frac{1}{2} kT_{0} R_{0}^{3} \int d^{3}
x n_{\nu}(x) \frac{v(x) - v(0)}{x} - R_{0}^{3} \int d^{3} x P_{\nu}(x),
\end{equation}

\begin{equation}
E = \frac{1}{2} \mu N_{\nu} - \frac{1}{2} kT_{0} R_{0}^{3} \int
d^{3}x n_{\nu}(x) \frac{v(x) + v(0)}{x} + R_{0}^{3} \int
d^{3} x \epsilon_{{\rm kin}} (x),
\end{equation}

\begin{equation}
S = \frac{1}{T} (E-F).
\end{equation}

We now turn to the numerical study of a system of selfgravitating
massive neutrinos with an arbitrarily chosen total mass $M = 10
M_{\odot}$, varying the cavity radius $R$. Owing to the scaling
properties, the system may be rescaled to any given mass.  For
definiteness, the $\nu_{\tau}$ mass is chosen as
$m_{\nu}$ = 17.2 keV$/c^{2}$ which is about the central value of the mass
region between 10 and 25 keV$/c^{2}$ \cite{RDV8}, that is interesting for our
scenario.  In Figure 4 we present our results for a gas of
neutrinos in a cavity of radius $R = 100 R_{0}$.  We find three
distinct solutions in the temperature interval $T = (0.049 \div 0.311)
T_{0}$; of these only two are physical solutions, namely, those for
which the free energy assumes a minimum.  The density distributions
corresponding to such two solutions are shown in the first plot
in$\,\,$
Figure 4.

The solution that can be continuously extended to any temperature
above the mentioned interval is referred to as `gas', whereas the
solution that continues to exist at low temperatures and
eventually becomes a degenerate Fermi gas at $T = 0$ is referred
to as `condensate'.  In Figure 4 we also plot
various extensive thermodynamical quantities (per neutrino) as functions
of the neutrino temperature.  The phase transition takes place at
a temperature $T_{t}$, where the free energy of the gas and of the
condensate become equal.  The transition temperature $T_{t} =
0.19442 T_{0}$ is indicated by the dotted line in the free-energy
plot.  The top dashed curve in the same plot corresponds to the
unphysical solution.  At $T = T_{t}$ the energy and the entropy
exhibit a discontinuity, and thus there will be a substantial
release of latent heat during the phase transition.  An
important and currently still open question is, how and to which
type of observable or unobservable matter or radiation this
latent heat, which can be interpreted as the binding energy of
the neutrino stars, will be transferred.

\newpage
\begin{figure}[p]
%\vspace{13cm}
\caption{Density distribution normalized to unity for
condensate-like (solid line) and gas-like (dashed line) solutions
at $T=T_{t}$.  Free energy, energy and entropy per particle
are shown as a
function of temperature.  Temperature, energy, and free energy
are in units of $T_{0}$.}
\end{figure}

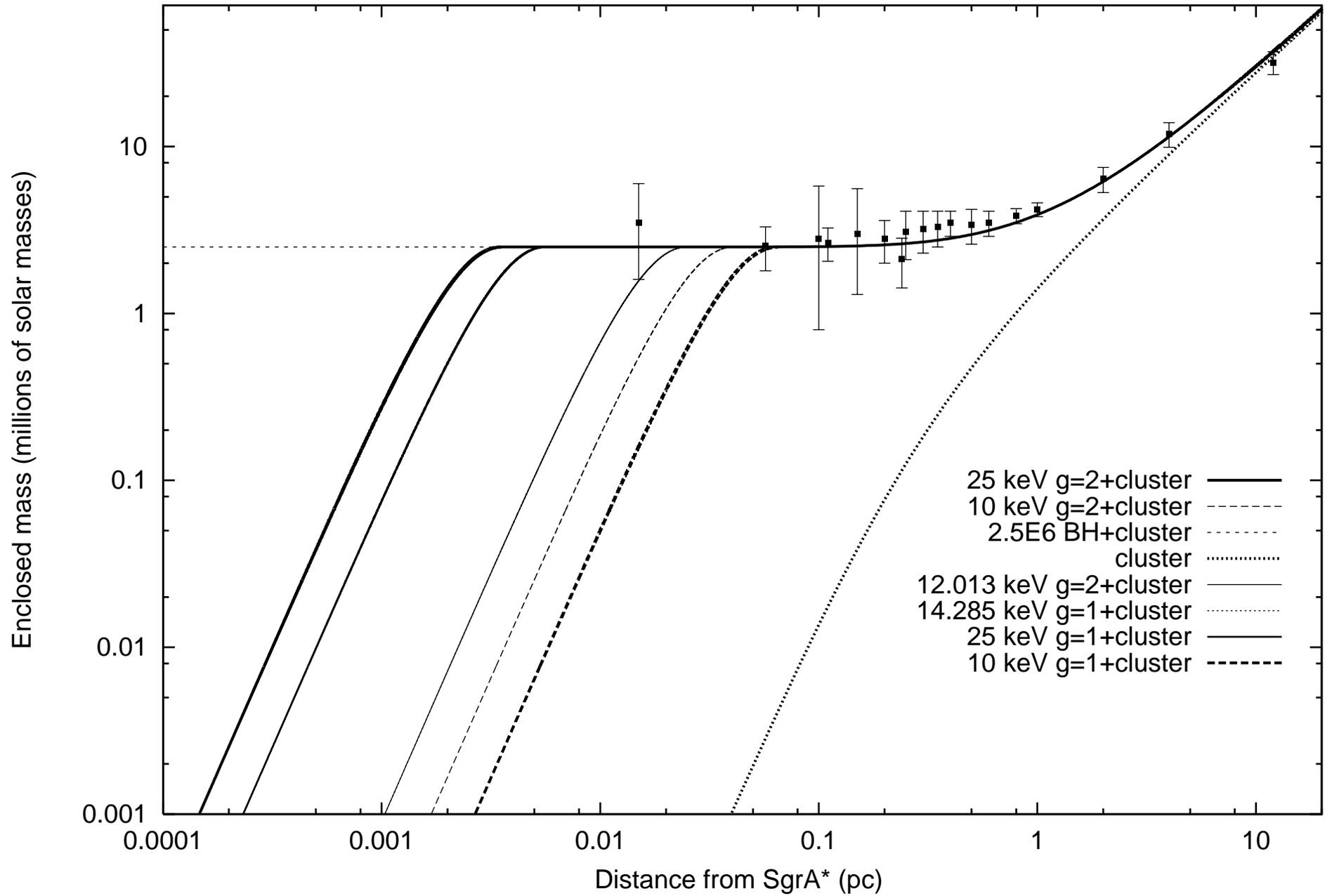
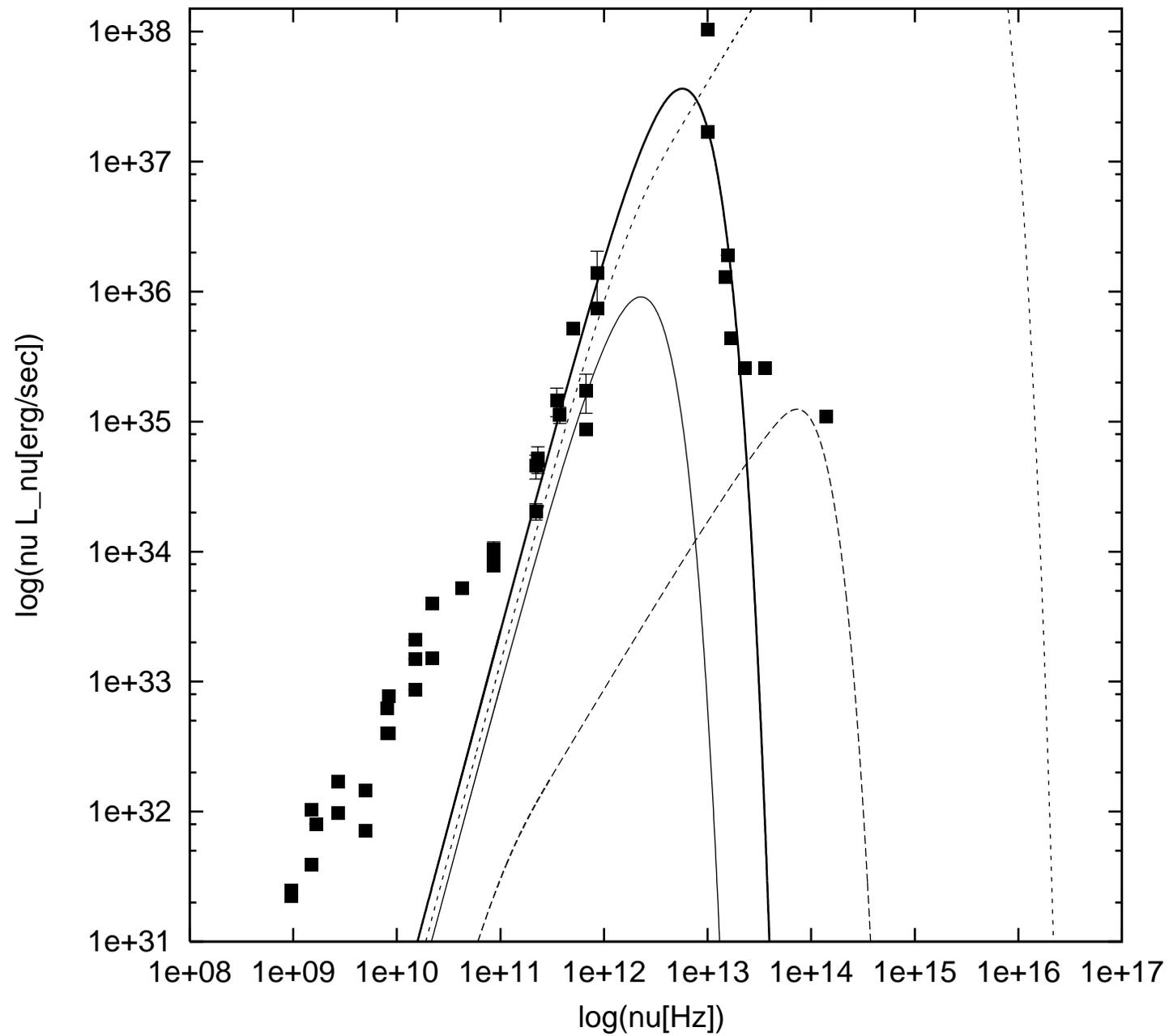
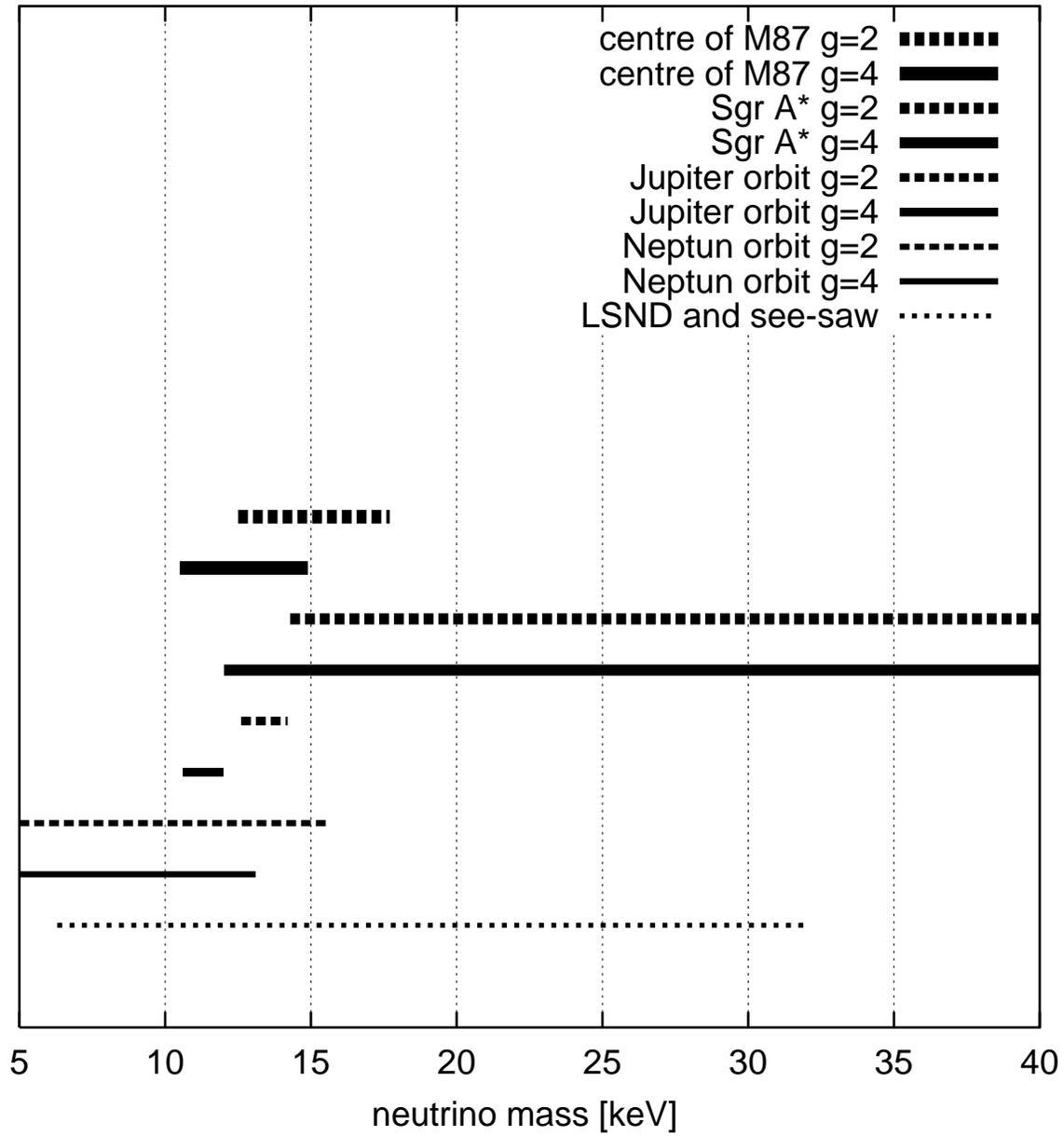
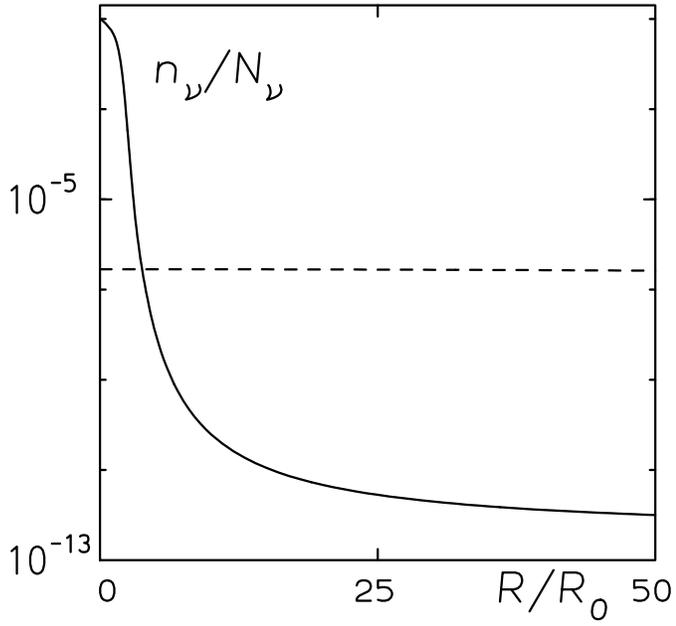
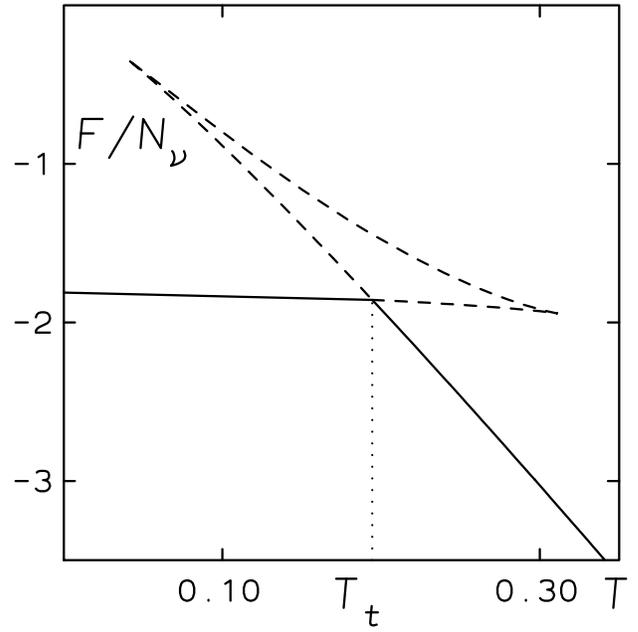
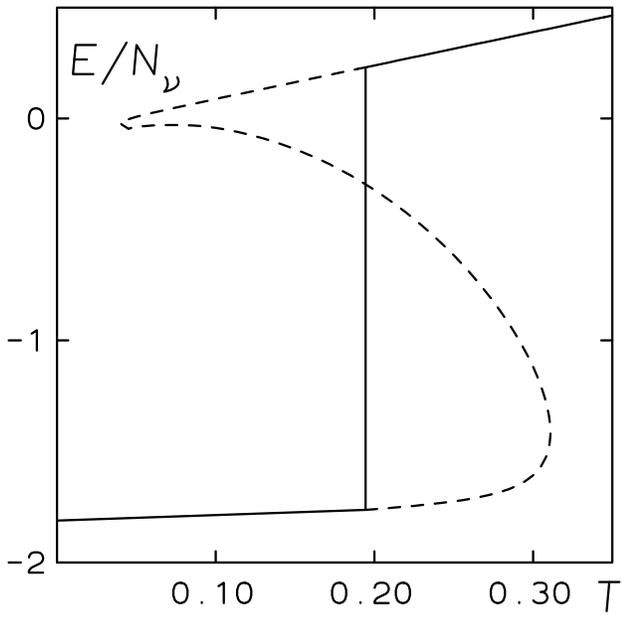
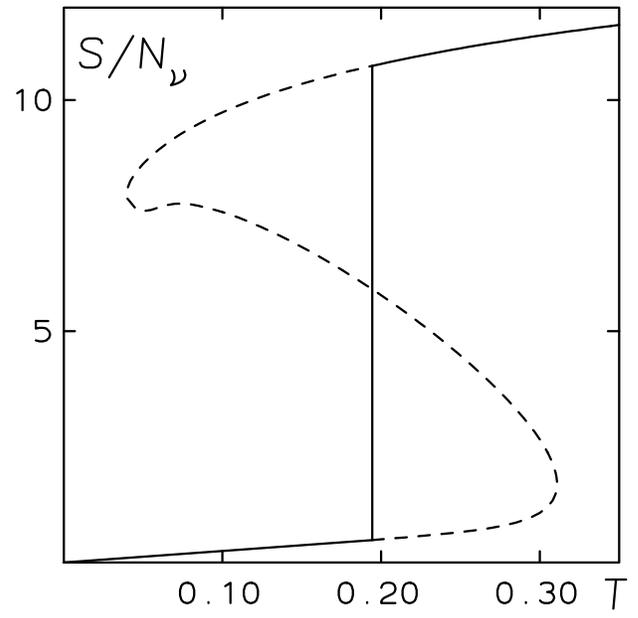
\end{document}